\documentclass[12pt]{article}
\usepackage{amsmath,amssymb,amsfonts,cite} 
\usepackage{graphicx}
\usepackage{stmaryrd}  
\usepackage{enumerate} 
\usepackage{stmaryrd}  
\usepackage{arydshln}  

\def\nn{\nonumber}

\def\gl{\mathfrak{gl}}
\def\Z{\mathbb{Z}}
\def\qdots{\mathinner{\mkern1mu\raise1pt\vbox{\kern7pt\hbox{.}}\mkern2mu \raise4pt\hbox{.}\mkern2mu\raise7pt\hbox{.}\mkern1mu}}

\begin{document}

\begin{center}
	{\large\bf The $\mathbb{Z}_2 \times \mathbb{Z}_2$-graded general linear
	Lie superalgebra
}\\
~~\\

{\large Phillip S. Isaac$^1$, N.I. Stoilova$^2$ and Joris Van der Jeugt$^3$}\\
~~\\
$^1$ 
School of Mathematics and Physics, The University of Queensland, St Lucia QLD 4072, Australia \\
$^2$ 
Institute for Nuclear Research and Nuclear Energy, 
Boul.\ Tsarigradsko Chaussee 72, 1784 Sofia, Bulgaria \\
$^3$ 
Department of Applied Mathematics, Computer Science and Statistics, Ghent University,
Krijgslaan 281-S9, B-9000 Gent, Belgium \\
~~\\
{\small Email: psi@maths.uq.edu.au, stoilova@inrne.bas.bg, Joris.VanderJeugt@UGent.be}
\end{center}

\begin{abstract}
We present a novel realisation of the $\mathbb{Z}_2\times\mathbb{Z}_2$-graded
	Lie superalgebra $\mathfrak{gl}(m_1,m_2|n_1,n_2)$ inside an algebraic
	extension of the enveloping algebra of the $\mathbb{Z}_2$-graded Lie
	superalgebra $\mathfrak{gl}(m|n)$, with $m=m_1+m_2$ and $n=n_1+n_2$. 
	A consequence of this realisation is
	that the representations of $\mathfrak{gl}(m|n)$ ``lift up'' to
	representations of $\mathfrak{gl}(m_1,m_2|n_1,n_2)$, with matrix
	elements differing only by a sign, which we are able to characterise
	concisely.
	
\end{abstract}

\section{Introduction}

Colour algebras or colour superalgebras are a generalization of Lie superalgebras introduced by Rittenberg and 
Wyler~\cite{RW1978v1,RW1978v2}, although related structures were discussed in the earlier work of Ree~\cite{Ree}.
Such algebras are characterized by a grading via an abelian group $\Gamma$, and the simplest case not coinciding with a Lie superalgebra is 
for $\Gamma=\Z_2\times\Z_2$.
Such $\Z_2\times\Z_2$ colour superalgebras have lately raised interest again, 
and in recent papers they are usually referred to as $\Z_2\times\Z_2$ graded Lie superalgebras.
We shall follow this terminology, although it is a bit misleading (as these algebras are not Lie superalgebras).

The notion of considering gradings beyond $\Z_2$ and their corresponding colour Lie (super)algebras goes back to the 1970's and 1980's.
Most of the attention on these structures has been on various mathematical aspects~\cite{Sch1983a,Sch1983b,Sylvestrov1997,CR2009,AS2017}. 
As remarked in~\cite{Ai2018}, physical applications of $\Z_2\times\Z_2$ graded Lie superalgebras are very limited 
compared with the fundamental importance of Lie algebras and Lie superalgebras in theoretical and mathematical physics. 
For some of the renewed physical interest, we refer to~\cite{Ai2018} and references therein.
Very recent work consists of the importance of $\Z_2\times\Z_2$ graded Lie superalgebras 
in the analysis of the L\'evy-Leblond equations~\cite{AKTT2016,AKTT2017}, and various
potential applications in the setting of supersymmetric and superconformal quantum mechanics
\cite{Bru,BruDup1,NaAmaDoi,Ueba,BruDup2}.

The algebra studied here is the $\Z_2\times\Z_2$ graded general linear Lie superalgebra $\mathfrak{gl}(m_1,m_2|n_1,n_2)$.
This algebra plays a special role among all finite dimensional $\Z_2\times\Z_2$ graded Lie superalgebras, because 
any other finite dimensional $\Z_2\times\Z_2$ graded Lie superalgebra (of appropriate
dimension) can be realised as a subalgebra of 
$\mathfrak{gl}(m_1,m_2|n_1,n_2)$~\cite{Sch1979,Tol2014}. As an example, we refer to the embedding of
the orthosymplectic $\Z_2\times\Z_2$ graded Lie superalgebra $\mathfrak{osp}(2m_1+1,2m_2|2n,0)$ in $\mathfrak{gl}(2m_1+1,2m_2|2n,0)$
used in~\cite{Tol2014}, where it is identified as one of the so-called para\-statistics algebras (see also~\cite{SV2018}).

In view of this, we considered it worthwhile to study $\mathfrak{gl}(m_1,m_2|n_1,n_2)$ and its representations.
Our main result is a rather straightforward realisation of $\mathfrak{gl}(m_1,m_2|n_1,n_2)$ inside an algebraic
extension of the enveloping algebra of the ordinary Lie superalgebra $\mathfrak{gl}(m|n)$, with $m=m_1+m_2$ and $n=n_1+n_2$. 

Our result should also been seen in the right context.
Already in~\cite{Sch1979}, Scheunert showed that there is a bijection between colour Lie (super)algebras 
(referred to as $\epsilon$ Lie algebras in~\cite{Sch1979}) and ordinary Lie (super)algebras
(provided the grading group $\Gamma$ is finitely generated).
This led to the perception that there is nothing beyond Lie algebras and Lie superalgebras.
Finding appropriate realisations of such bijections is, however, not trivial.

Many years after Scheunert's result, McAnally and Bracken~\cite{MB1997}
developed a technique to construct the Lie superalgebra $\gl(p|q)$ out of the colour Lie algebra $\gl(n)$ for $p+q=n$.
This ``colour $\gl(n)$'' is a colour Lie algebra whose $n^2$ basis elements $E_{ij}$ are graded according to a general grading
group $\Gamma$, with a bracket depending on a phase function for $\Gamma$.
The technique of~\cite{MB1997} is inspired by the so-called Klein operators discussed by Kleeman~\cite{Kleeman1983,Kleeman1985}.
In~\cite{MB1997}, the authors give a realisation of the $\gl(p|q)$ generators as elements of the enveloping
algebra of colour $\gl(n)$, thus establishing the mentioned bijection.
Although this result is remarkable, the construction itself is rather complicated, even if one tries to apply this to
a simple case like $\gl(m_1,m_2|n_1,n_2)$.

The present contribution is closely related to the above construction, but we follow a
path in the opposite direction.
Our starting point is the Lie superalgebra $\gl(m|n)$, and inside an algebraic extension
of its enveloping algebra we construct operators that
realise $\gl(m_1,m_2|n_1,n_2)$.
Our construction is so straightforward that we need in the end only two additional
multiplicative operators in order to get the 
$\gl(m_1,m_2|n_1,n_2)$ generators out of those from $\gl(m|n)$.

Having established the realisation of $\gl(m_1,m_2|n_1,n_2)$ inside an algebraic exension of the
enveloping algebra of $\gl(m_1+m_2|n_1+n_2)$, we demonstrate how the result impacts the
representation theory. Specifically, we show how our algebraic results allow us to lift up
irreducible covariant representations of $\gl(m|n)$ to representations of the colour algebra
$\gl(m_1,m_2|n_1,n_2)$. The addition of the afore-mentioned multiplicative operators allow us to
easily and explicitly understand how the matrix elements are affected only by a sign.


\section{Definitions and algebraic structure}

It is instructive to briefly recall the definition of a $\mathbb{Z}_2$-graded Lie superalgebra
\cite{SNR1976,Kac1977}. 
Let $\mathfrak{g}$ be a $\mathbb{Z}_2$-graded vector space over the complex field
$\mathbb{C}$, in the sense that it decomposes as a direct sum 
$$
\mathfrak{g} = \mathfrak{g}_{0} \oplus \mathfrak{g}_{1}.
$$
For a homogeneous element $x\in \mathfrak{g}_0$ or $\mathfrak{g}_1$, it is also
useful to introduce the {\em degree} as 
$$
d(x) =\left\{ 
\begin{array}{rl}
0; & x\in \mathfrak{g}_{0},\\
1; & x\in \mathfrak{g}_{1}.
\end{array} \right.
$$
Then $\mathfrak{g}$ is a Lie superalgebra if it is endowed with 
a bilinear product $\llbracket \cdot,\cdot \rrbracket:
\mathfrak{g}\otimes \mathfrak{g}\rightarrow \mathfrak{g}$ that preserves
grading, i.e. if $x\in \mathfrak{g}_{d(x)}$, $y\in \mathfrak{g}_{d(y)}$, then $\llbracket x,y\rrbracket\in \mathfrak{g}_{d(x)+d(y)},$ with the sum $d(x)+d(y)$ taken modulo
2, and products involving nonhomogeneous elements are determined by extending
the Lie product $\llbracket \cdot,\cdot \rrbracket$ through linearity. The Lie product must also
satisfy
\begin{align}
	\llbracket x,y \rrbracket & = -(-1)^{d(x)\cdot d(y)} \llbracket y,x
	\rrbracket, \mbox{ (graded antisymmetry)} \label{asym}\\ 
	\llbracket \llbracket x, y \rrbracket, z \rrbracket 
	& = \llbracket x,\llbracket y,z\rrbracket \rrbracket - (-1)^{d(x)\cdot
	d(y)} \llbracket y,\llbracket x,z\rrbracket \rrbracket, \mbox{ (graded
	Jacobi identity).} \label{jacobi}
\end{align}

An important example for our purposes is the Lie superalgebra $\mathfrak{gl}(m|n)$.
The standard basis of $\mathfrak{gl}(m|n)$ is given by the
(homogeneous) elementary matrices $E_{ij}$ (with 1 in the entry of row $i$, column $j$ and 0
elsewhere) with grading conveniently characterised via the indices. To this end, set
$$
d_i = \left\{   
\begin{array}{rl}
0; & i=1,\ldots,m, \\
1; & i=m+1,\ldots,m+n.
\end{array}
\right.
$$
We then define $d_{ij} := d(E_{ij}) = d_i + d_j$ modulo 2. 

The $\mathbb{Z}_2$-graded Lie product is then given by
\begin{equation}
\llbracket E_{ij},E_{k\ell}\rrbracket 
= 
\delta_{jk} E_{i\ell} - (-1)^{d_{ij}\cdot d_{k\ell}} \delta_{i\ell} E_{kj}.
	\label{LieProd}
\end{equation}
The universal enveloping algebra $U(\mathfrak{gl}(m|n))$ is then the unital associative
algebra with the free product of generators $E_{ij}$ subject to the algebraic relations
\begin{equation}
E_{ij}E_{k\ell} - (-1)^{d_{ij}\cdot d_{k\ell}} E_{k\ell} E_{ij} = 
\delta_{jk} E_{i\ell} - (-1)^{d_{ij}\cdot d_{k\ell}} \delta_{i\ell} E_{kj}.
\label{LieProdU}
\end{equation}

Having introduced $\mathfrak{gl}(m|n)$ in this standard way, the definition of its
various extensions to the 
$\mathbb{Z}_2\times \mathbb{Z}_2$-graded Lie superalgebras
$\mathfrak{gl}(m_1,m_2|n_1,n_2)$ generalises naturally. Following Rittenberg
and Wyler \cite{RW1978v1,RW1978v2} and Tolstoy \cite{Tol2014}, 
let $\tilde{\mathfrak{g}}$ be a $\mathbb{Z}_2\times \mathbb{Z}_2$-graded vector space over the complex field
$\mathbb{C}$, that decomposes as
$$
\tilde{\mathfrak{g}} = \tilde{\mathfrak{g}}_{(0,0)} \oplus
\tilde{\mathfrak{g}}_{(1,1)} \oplus \tilde{\mathfrak{g}}_{(1,0)} \oplus
\tilde{\mathfrak{g}}_{(0,1)}.
$$
For a homogeneous element $x_a\in \tilde{\mathfrak{g}}_a$, $a=(0,0),$ $(1,1)$, $(1,0)$,
or $(0,1)$, define the {\em degree} as 
$\tilde{d}(x_a) = a$.
Then $\tilde{\mathfrak{g}}$ is a $\mathbb{Z}_2\times \mathbb{Z}_2$-graded Lie superalgebra if it is endowed with 
a bilinear product $\llbracket \cdot,\cdot \rrbracket:
\tilde{\mathfrak{g}}\otimes \tilde{\mathfrak{g}}\rightarrow
\tilde{\mathfrak{g}}$ that preserves this
grading, i.e. if $x\in \tilde{\mathfrak{g}}_{\tilde{d}(x)}$, $y\in
\tilde{\mathfrak{g}}_{\tilde{d}(y)}$, then $\llbracket x,y\rrbracket\in
\tilde{\mathfrak{g}}_{\tilde{d}(x)+\tilde{d}(y)}.$ In this case, the sum is
componentwise addition modulo 2. That is, if $\tilde{d}(x) = (x_1,x_2)$ and
$\tilde{d}(y) = (y_1,y_2)$ then $\tilde{d}(x)+\tilde{d}(y) = (x_1+y_1,x_2+y_2)$
with the sum in each component being taken modulo 2.
As before, products involving nonhomogeneous elements are determined by extending
$\llbracket \cdot,\cdot \rrbracket$ through linearity. 

The product must also satisfy
\begin{align}
	\llbracket x,y \rrbracket & = -(-1)^{\tilde{d}(x)\cdot \tilde{d}(y)} \llbracket y,x
	\rrbracket, \mbox{ (graded antisymmetry)} \label{asym2}\\ 
	\llbracket \llbracket x, y \rrbracket, z \rrbracket 
	& = \llbracket x,\llbracket y,z\rrbracket \rrbracket - (-1)^{\tilde{d}(x)\cdot
	\tilde{d}(y)} \llbracket y,\llbracket x,z\rrbracket \rrbracket, \mbox{ (graded
	Jacobi identity).} \label{jacobi2}
\end{align}
In the case that $\tilde{d}(x)=(x_1,x_2)$ and
$\tilde{d}(y)=(y_1,y_2)$ then the grading factors appearing in the above
expressions are determined by the ``dot product'' $\tilde{d}(x)\cdot \tilde{d}(y) = x_1y_1 +
x_2y_2$ modulo 2.

The main object of our study in this paper is the
$\mathbb{Z}_2\times\mathbb{Z}_2$-graded general linear superalgebra
$\mathfrak{gl}(m_1,m_2|n_1,n_2)$, where the labels $m_1$, $m_2$,
$n_1$ and $n_2$ are non-negative integers. 
As in the $\mathbb{Z}_2$-graded case, a convenient basis is the standard one
comprising (homogeneous) elementary matrices $\tilde{E}_{ij}$ with 1 in the
entry of row $i$, column $j$ and 0 elsewhere. To describe the grading of these
basis elements, we use the graded index notation
introduced in \cite{Tol2014}, whereby
\begin{align*}
\tilde{d}_i &= 
\left\{  
\begin{array}{rl}
	(0,0); & i=1,\ldots,m_1\\
	(1,1); & i=m_1+1,\ldots,m_1+m_2\\
	(1,0); & i=m_1+m_2+1,\ldots,m_1+m_2+n_1\\
	(0,1); & i=m_1+m_2+n_1+1,\ldots,m_1+m_2+n_1+n_2,
\end{array}
\right.
\end{align*}
where we adopt the convention that in a situation where one of the labels $m_1$, $m_2$,
$n_1$ or $n_2$ is zero, the case $i=j,\ldots,j-1$ does not exist. \footnote{Note that in the
case where two or more of these labels are zero, the algebra reduces to a Lie superalgebra
or Lie algebra.}
We then define $\tilde{d}_{ij}:=\tilde{d}(\tilde{E}_{ij}) =
\tilde{d}_i+\tilde{d}_j$, where the sum is taken componentwise, modulo 2, as
described earlier. The $\mathbb{Z}_2\times\mathbb{Z}_2$-graded Lie product then
takes on the same form as (\ref{LieProd}), namely
\begin{equation}
\llbracket \tilde{E}_{ij},\tilde{E}_{k\ell}\rrbracket 
= 
\delta_{jk} \tilde{E}_{i\ell} - (-1)^{\tilde{d}_{ij}\cdot \tilde{d}_{k\ell}} \delta_{i\ell}
\tilde{E}_{kj}.
\label{tildeLieProd}
\end{equation}
The universal enveloping algebra $U(\mathfrak{gl}(m_1,m_2|n_1,n_2))$ is then the unital associative
algebra with the free product of generators $\tilde{E}_{ij}$ subject to the algebraic relations
\begin{equation}
\tilde{E}_{ij}\tilde{E}_{k\ell} - (-1)^{\tilde{d}_{ij}\cdot \tilde{d}_{k\ell}}
\tilde{E}_{k\ell} \tilde{E}_{ij} = 
\delta_{jk} \tilde{E}_{i\ell} - (-1)^{\tilde{d}_{ij}\cdot \tilde{d}_{k\ell}} \delta_{i\ell}
\tilde{E}_{kj}.
\label{tildeLieProdU}
\end{equation}

It is worth remarking that the universal enveloping algebras of both the $\mathbb{Z}_2$ and $\mathbb{Z}_2\times
\mathbb{Z}_2$ graded superalgebras described above enjoy relations with both
commutators and anticommutators. A worthwhile exercise is to compare the
$\mathbb{Z}_2\times \mathbb{Z}_2$-graded $\mathfrak{gl}(m_1,m_2|n_1,n_2)$ with
the $\mathbb{Z}_2$-graded $\mathfrak{gl}(m_1+m_2|n_1+n_2)$.  

For readers unfamiliar with the topic, it should once again be emphasized that a $\mathbb{Z}_2\times
\mathbb{Z}_2$ graded Lie superalgebra is not a Lie superalgebra nor a Lie algebra, but a different structure.
For example, comparing $\gl(1,1|1,1)$ and $\gl(2|2)$ commutators can become anticommutators and vice versa.
E.g.\ in $U(\gl(2|2))$ the bracket of $E_{13}$ and $E_{41}$ is an anticommutator $\{ E_{13},E_{41} \}=E_{43}$, whereas
in $U(\gl(1,1|1,1))$ it is a commutator: $[ \tilde E_{13},\tilde E_{41} ]=-\tilde E_{43}$.
Similarly, the commutator $[ E_{21},E_{13} ]=E_{23}$ in $\gl(2|2)$ becomes an anticommutator 
$\{ \tilde E_{21},\tilde E_{13} \}=\tilde E_{23}$ in $\gl(1,1|1,1)$.


\section{Algebraic extension of $U(\mathfrak{gl}(m|n))$}

The goal of this section is to give a realisation of
$U(\mathfrak{gl}(m_1,m_2|n_1,n_2))$ in terms of an algebraic extension of
$U(\mathfrak{gl}(m|n))$. This ultimately allows us to establish an injection mapping
irreducible $\mathfrak{gl}(m|n)$-modules into irreducible
$\mathfrak{gl}(m_1,m_2|n_1,n_2)$-modules.

First of all, define the following elements $H_k$ from $U(\mathfrak{gl}(m|n))$, which are just sums of $E_{ii}$'s:
\begin{equation}
H_{k}=E_{11}+E_{22}+\cdots +E_{kk}.
\end{equation}
Then the following relations hold in $U(\mathfrak{gl}(m|n))$
\begin{align}
	&[ H_k, E_{ij}] = 0, \mbox{\quad for \quad } k<i<j;\;\;i<j\leq k; \;\; i>j>k; \;\; k\geq i>j,  \nn \\ 
	& [ H_k, E_{ij}] = E_{ij}, \mbox{\quad for \quad } i\leq k<j;\nn\\
	& [ H_k, E_{ij}] = -E_{ij}, \mbox{\quad for \quad } i>k\geq j. \label{relEk}
\end{align}
The operators we actually need are the following elements:
\begin{equation}
B_{k}=(-1)^{H_{k}}; \quad B_{k}^{-1}=(-1)^{-H_{k}}.
\label{Bk}
\end{equation}
Such elements can be viewed as belonging to an extension or closure of the universal enveloping algebra of $\mathfrak{gl}(m|n)$,
using 
\[
(-1)^A=\exp(i\pi A)=\sum_j \frac{(i \pi A)^j}{j!}.
\]
We denote this algebraic extension of $U(\mathfrak{gl}(m|n))$ that includes such
elements by $\tilde{U}(m,n)$.
Of course, if one considers finite-dimensional $\gl(m|n)$ representations in which the action of the Cartan generators is diagonal,
the interpretation of elements such as $(-1)^{H_k}$ in $\tilde{U}(m,n)$ is uncomplicated.

Utilising the commutation relations~\eqref{relEk}, it is rather straightforward to deduce the following relations between
the $B_k$'s and the $E_{ij}$'s:
\begin{align}
	& B_kE_{ij} = E_{ij}B_k, \mbox{\quad for \quad } k<i<j;\;\;i<j\leq k; \;\; i>j>k; \;\; k\geq i>j,  \nn \\ 
	& B_kE_{ij} = -E_{ij}B_k, \mbox{\quad for \quad } i\leq k<j;\;\; i>k\geq j, \label{relBk} \\
	& B_k^{-1}E_{ij} = E_{ij}B_k^{-1}, \mbox{\quad for \quad } k<i<j;\;\;i<j\leq k; \;\; i>j>k; \;\; k\geq i>j,  \nn \\ 
	& B_k^{-1}E_{ij} = -E_{ij}B_k^{-1}, \mbox{\quad for \quad } i\leq k<j;\;\; i>k\geq j. \label{relBk-1}
\end{align}
In other words, the plus or minus sign when switching $B_k$ with $E_{ij}$ depends on whether $k$ sits inside or outside the range 
determined by $i$ and $j$.

We are now in a position to describe the main result of the paper, namely a realisation of $\gl(m_1,m_2|n_1,n_2)$
in $\tilde{U}(m|n)$, for $m=m_1+m_2$ and $n=n_1+n_2$.
For this, all we need is to multiply certain $\gl(m|n)$ generators $E_{ij}$ by $B_{m_1}$ and/or $B_m$.
\begin{align}
{\rm (a)\ } & \tilde{E}_{ij} = E_{ij},  \mbox{\;for\;} i,j=1,\cdots ,m;\;\;i,j=m+1,\cdots, m+n_1;\nn\\
  &\qquad \qquad \;\; \mbox{and\;} i,j=m+n_1+1,\cdots, m+n,  \nn \\ 
{\rm (b)\ }	&\tilde{E}_{ij} = E_{ij}B_{m_1},  \mbox{\;for\;} i=1,\cdots, m;\; 
	j=m+1,\cdots, m+n_1,  \nn \\
{\rm (c)\ }	& \tilde{E}_{ij} = E_{ij}B_{m},  \mbox{\;for\;} i=m+1,\cdots, m+n_1;
	 \; j=m+n_1+1,\cdots, m+n,  \nn \\
{\rm (d)\ }	& \tilde{E}_{ij} = E_{ij}B_{m_1}B_{m},  \mbox{\;for\;} i=1,\cdots,m;
	j=m+n_1+1,\cdots, m+n,  \nn \\
{\rm (e)\ }	 &\tilde{E}_{ij} = B_{m_1}^{-1}E_{ij},  \mbox{\;for\;} i=m+1,\cdots, m+n_1;
	j=1,\cdots, m,  \nn \\
{\rm (f)\ }	& \tilde{E}_{ij} = B_{m}^{-1}E_{ij},  \mbox{\;for\;} i=m+n_1+1,\cdots, m+n;\ 
	  j=m+1,\cdots, m+n_1,  \nn \\
{\rm (g)\ }	 &\tilde{E}_{ij} = B_{m_1}^{-1}B_{m}^{-1}E_{ij},  \mbox{\;for\;} i=m+n_1+1,\cdots, m+n;
	j=1,\cdots,m. \label{trelBk-1}
\end{align}
So simply said, the index set $[1,m+n]\times[1,m+n]$ is split into 7 regions: region (a) where there
is no difference between $E_{ij}$ and $\tilde E_{ij}$; regions (b), (c) and (d) where a multiplication to the right takes place 
by $B_{m_1}$, $B_m$ or their product; and regions (e), (f) and (g) where a multiplication to the left takes place 
by $B_{m_1}^{-1}$, $B_m^{-1}$ or their product. It should be noted that in the case
$m_1=0$, then $B_{m_1}=I$.

It is straightforward to check that equations~\eqref{trelBk-1} give a realisation of $U(\gl(m_1,m_2|n_1,n_2))$ 
inside $\tilde{U}(m,n)$. For this purpose we must check 
that relations~\eqref{tildeLieProdU} hold. 
It is quite a task to check all possible cases from all possible index regions, and it would be 
too tedious to write down all of these.
We restrict ourselves to giving just two representative cases, where moreover some sign change takes place.
As a first example, let 
$i=1,\cdots, m_1$, $j=m+1,\cdots,m+n_1$, $k=m+1,\cdots, m+n_1$, $l=m_1+1,\cdots, m$. Then
$\tilde{E}_{ij}\in \gl(m_1,m_2|n_1,n_2)_{(1,0)}$, $\tilde{E}_{kl}\in \gl(m_1,m_2|n_1,n_2)_{(0,1)}$ and 
 $\tilde{d}_{ij}.\tilde{d}_{kl}=0$. So
\begin{align}
&\tilde{E}_{ij}\tilde{E}_{kl}-\tilde{E}_{kl}\tilde{E}_{ij}=E_{ij}B_{m_1}B_{m_1}^{-1}E_{kl}-
B_{m_1}^{-1}E_{kl}E_{ij}B_{m_1}\nn\\
& =E_{ij}E_{kl}+E_{kl}E_{ij}=\delta_{jk}E_{il}=\delta_{jk}\tilde{E}_{il}.
\end{align}
The sign change is because for the chosen index sets $E_{ij}B_{m_1}=-B_{m_1}E_{ij}$,
and is necessary because $E_{ij}$ and $E_{kl}$ belong to $\gl(m|n)_1$.
As a second example, let 
$i=1,\cdots, m_1$, $j=m+1,\cdots,m+n_1$, $k=m+n_1+1,\cdots, m+n$, $l=1,\cdots, m_1$. Then
again $\tilde{E}_{ij}\in \gl(m_1,m_2|n_1,n_2)_{(1,0)}$, $\tilde{E}_{kl}\in \gl(m_1,m_2|n_1,n_2)_{(0,1)}$ and 
 $\tilde{d}_{ij}.\tilde{d}_{kl}=0$. So
\begin{align}
&\tilde{E}_{ij}\tilde{E}_{kl}-\tilde{E}_{kl}\tilde{E}_{ij}=E_{ij}B_{m_1}B_{m_1}^{-1}B_{m}^{-1}E_{kl}-
B_{m_1}^{-1}B_{m}^{-1}E_{kl}E_{ij}B_{m_1}\nn\\
&=-B_{m}^{-1}E_{ij}E_{kl}-B_{m}^{-1}E_{kl}E_{ij}
=-B_{m}^{-1}(E_{ij}E_{kl}+E_{kl}E_{ij})\nn\\
&=-B_{m}^{-1}\delta_{il}E_{kj}=-\delta_{il}\tilde{E}_{kj}.
\end{align}
It is worth pointing out that these results are also valid for the cases where one of the
labels $m_1$, $m_2$, $n_1$, $n_2$ is zero. 

Perhaps more interesting than the actual proof is how we found this solution.
It relies on comparing the brackets for $\tilde E_{ij}$ ($1\leq i,j \leq k$)
with those of $E_{ij}$ ($1\leq i,j \leq k$), for $k=1,2,\ldots$, and gradually increase $k$.
For $k\leq m_1+m_2=m$, the commutators are the same, and nothing should change.
For $k=m+1$, one sees that for the first time certain commutators between $E_{ik}$ and $E_{jl}$
($j,l<k$) should become anticommutators for the corresponding elements $\tilde E_{ik}$ and $\tilde E_{jl}$.
A close inspection shows that it is sufficient to multiply the elements $E_{i,m+1}$ by $B_{m_1}$, 
as $m_1+1$ is the first index for which the elements $\tilde E_{jl}$ have grading~$(1,1)$.
One can continue this process with a similar outcome, until $k=m+n_1$.
For $k=m+n_1+1$ a change from commutator of $E_{m+n_1,m+n_1+1}$ and $E_{m,m+n_1}$ to anticommutator should take place,
and careful inspection leads to a multiplication by $B_{m}$ to achieve this.
Then one continues increasing $k$, always examining the mutual relations, until one reaches the value $m+n$.
 

\section{Characterisation of representations}

To see how easy it is to lift the $\gl(m_1,m_2|n_1,n_2)$ realisation to representations, 
let us restrict to the important class of covariant representations of $\gl(m|n)$. Given
the universal nature of our construction, it would be possible to apply it to the class of
irreducible unitary representations \cite{GZ1990}, but for the purposes of this article,
it is sufficient to demonstrate our point on the covariant representations.
Such representations are labelled by a partition $\lambda$ satisfying the $(m,n)$-hook condition,
and have a highest weight with coordinates $(\mu_{1r},\mu_{2r},\ldots,\mu_{rr})$ (where $r=m+n$) in
the common weight basis $(\epsilon_1,\ldots,\epsilon_m,\delta_1,\ldots,\delta_n)$ of $\gl(m|n)$
(see~\cite{SV2010} for a detailed description).
The basis vectors of such a representation are labelled by particular triangular Gelfand-Zetlin patterns,
of the following form:
\begin{equation}
|\mu) = \left|
\begin{array}{lclllcll}
\mu_{1r} & \cdots & \mu_{m-1,r} & \mu_{mr} & \mu_{m+1,r} & \cdots & \mu_{r-1,r}
& \mu_{rr}\\
\mu_{1,r-1} & \cdots & \mu_{m-1,r-1} & \mu_{m,r-1} & \mu_{m+1,r-1} & \cdots
& \mu_{r-1,r-1} & \\
\vdots & \vdots &\vdots &\vdots & \vdots & \qdots & & \\
\mu_{1,m+1} & \cdots & \mu_{m-1,m+1} & \mu_{m,m+1} & \mu_{m+1,m+1} & & & \\
\mu_{1m} & \cdots & \mu_{m-1,m} & \mu_{mm} & & & & \\
\mu_{1,m-1} & \cdots & \mu_{m-1,m-1} & & & & & \\
\vdots & \qdots & & & & & & \\
\mu_{11} & & & & & & &
\end{array}
\right),
\label{mu}
\end{equation}
where all entries $\mu_{ij}$ are positive integers.
For a certain representation, the top row is fixed and corresponds to the highest weight.
The other labels $\mu_{ij}$ should satisfy a number of conditions, described in~\cite[Proposition~6]{SV2010}.

For convenience, let us also introduce a notation for the sum of all entries on row~$i$ of~\eqref{mu}:
\[
| \mu_i| = \mu_{1,i}+\mu_{2,i}+\cdots+\mu_{i,i}.
\]
From~\cite[Theorem~7]{SV2010}, one gets the following action of the generators of the Cartan subalgebra of $\gl(m|n)$:
\begin{equation}
E_{ii} |\mu) = (|\mu_i| - |\mu_{i-1}|) \; |\mu),
\end{equation}
where $|\mu_0|=0$.
As a consequence, the operators $B_k$ act as
\begin{equation}
B_k |\mu) = (-1)^{|\mu_k|} |\mu).
\end{equation}

The main result of~\cite{SV2010} was the explicit computation of all matrix elements of the $gl(m|n)$ Chevalley generators in 
such a covariant representation, i.e.\ of
\[
(\mu'| E_{i,i+1} |\mu) \hbox{ and }(\mu'| E_{i+1,i} |\mu) \qquad i=1,2,\ldots,r-1.
\]

Following the construction of the previous section, the covariant representations described here augment to irreducible 
representations of $\gl(m_1,m_2|n_1,n_2)$, with the following diagonal actions
\begin{equation}
\tilde E_{ii} |\mu) = (|\mu_i| - |\mu_{i-1}|) \; |\mu),
\end{equation}
and with 
\begin{align*}
&(\mu'| \tilde E_{i,i+1} |\mu)= (\mu'| E_{i,i+1} |\mu) \qquad i\ne m, m+n_1,\\
&(\mu'| \tilde E_{m,m+1} |\mu)= (-1)^{|\mu_{m_1}|} (\mu'| E_{m,m+1} |\mu) ,\\
&(\mu'| \tilde E_{m+n_1,m+n_1+1} |\mu)= (-1)^{|\mu_{m}|} (\mu'| E_{m+n_1,m+n_1+1} |\mu) ,\\
&(\mu'| \tilde E_{i+1,i} |\mu)= (\mu'| E_{i+1,i} |\mu) \qquad i\ne m, m+n_1,\\
&(\mu'| \tilde E_{m+1,m} |\mu)= (-1)^{|\mu_{m_1}|} (\mu'| E_{m+1,m} |\mu) ,\\
&(\mu'| \tilde E_{m+n_1+1,m+n_1} |\mu)= (-1)^{|\mu_{m}|} (\mu'| E_{m+n_1+1,m+n_1} |\mu) .
\end{align*}
Matrix elements of other basis elements $\tilde E_{ij}$ follow, as usual, by taking appropriate $\gl(m_1,m_2|n_1,n_2)$ brackets.

It should not be too surprising that the matrix elements of these covariant
representations of $\mathfrak{gl}(m_1,m_2|n_1,n_2)$ essentially differ to the matrix elements of the
Lie superalgebra $\mathfrak{gl}(m_1+m_2|n_1+n_2)$ by a sign in
certain entries. Firstly, it is easy to see that the quadratic Casimir of $\mathfrak{gl}(m_1,m_2|n_1,n_2)$ has
the same form as that of $\mathfrak{gl}(m|n)$:
$$
I_2 = \sum_{i,j}(-1)^{\tilde{d}_{ij}\cdot\tilde{d}_{ij}}\tilde{E}_{ij}\tilde{E}_{ji} .
$$
Then, on a highest weight representation with highest weight
$(\mu_{1r},\ldots,\mu_{mr},\mu_{m+1,r},\ldots,\mu_{rr})$, the eigenvalues also have the same form:
$$
\sum_{i=1}^m(\mu_{ir}(\mu_{ir}+m-n-2i+1)) -
\sum_{k=1}^n(\mu_{m+k,r}(\mu_{m+k,r}+m+n-2k+1)).
$$
Using characteristic identity techniques \cite{GJ1983,GIW2013,GIW2014}, the resulting
formulae for the square of the matrix elements of the generators for
$\mathfrak{gl}(m_1,m_2|n_1,n_2)$ and $\mathfrak{gl}(m_1+m_2|n_1+n_2)$ are the same, and
therefore the matrix entries on such a representation can only differ by a sign.

\section{Conclusion and outlook}

We have provided a straightforward and explicit construction of the $\mathbb{Z}_2\times\mathbb{Z}_2$-graded
colour Lie superalgebra $\mathfrak{gl}(m_1,m_2|n_1,n_2)$ inside an algebraic extension of the
enveloping algebra of the Lie superalgebra $\mathfrak{gl}(m_1+m_2|n_1+n_2)$. The
motivation for studying the general linear $\mathfrak{gl}(m_1,m_2|n_1,n_2)$ is the fact that Ado's Theorem
is known to also hold for the colour case \cite{Sch1979,Tol2014}, so some knowledge of
explicit results should be worthwhile to develop. The simplicity of our
construction allows us to lift up representations of $\mathfrak{gl}(m|n)$ to representations of
$\mathfrak{gl}(m_1,m_2|n_1,n_2)$, where $m_1+m_2=m$ and $n_1+n_2=n$. We furthermore
demonstrated this process on the class of irreducible covariant tensor representations.

There are various immediate consequences of the correspondence presented in this article. For instance, using equations
(\ref{trelBk-1}), it is possible to write down Serre relations \cite{Zh2014} satisfied by the Chevalley
generators $\tilde E_{i,i+1}$ and $\tilde E_{i+1,i}$ of $\mathfrak{gl}(m_1,m_2|n_1,n_2)$.
In some cases, the relative signs of the coefficients in the relations may change as a
result of equations (\ref{relBk}) and (\ref{relBk-1}), which is particularly evident in the
cases where one of the labels $m_1$, $m_2$, $n_1$ or $n_2$ is zero. Such a presentation
would allow, for example, one to develop the quantum analogue and express the relations of the associated quantum group
along the lines of \cite{FLV1991,KT1991,Sch1993}.

\section*{Acknowledgements}

P.S.\ Isaac would like to thank the Department of Applied Mathematics, Computer Science and
Statistics at Ghent University for its warm hospitality, where this work was initiated.
N.I.\ Stoilova was supported by the Bulgarian National Science Fund, grant KP-06-N28/6, 
and J.\ Van der Jeugt was supported by the EOS Research Project 30889451.

\end{document}